# VOID INDUCED MOLECULE $C_{23}H_{12}^{++}$ COULD REPRODUCE THE INFRARED SPECTRUM (3 TO 20μm) OF INTERSTELLAR GAS AND DUST

Norio Ota

Graduate School of Pure and Applied Sciences, University of Tsukuba, *1-1-1 Tenoudai Tsukuba-city 305-8571, Japan*

In order to find out a selected number of molecules to reproduce the infrared spectrum of interstellar gas and dust, model coronene molecules with void and charge have been computed using density functional theory. Among them, a single void induced cation $C_{23}H_{12}^{++}$ have successfully reproduced a wide range spectrum (3~20 μm) of typical interstellar gas and dust. Well known astronomically observed emission peaks were 3.3, 6.2, 7.6, 7.8, 8.6, 11.2, 12.7, 13.5 and 14.3 μm. Whereas, calculated peaks of $C_{23}H_{12}^{++}$ were 3.2, 6.4, 7.6, 7.8, 8.6, 11.4, 12.9, 13.5, and 14.4 μm. It should be noted that a single kind of molecule could reproduce very well not depending on the decomposition method using many polycyclic aromatic hydrocarbon (PAH) data. Such coincidence suggested that some astronomical chemical evolution may select a particular PAH. Molecular structure of $C_{23}H_{12}^{++}$ was dramatically deformed by the Jahn-Teller effect. There is a featured carbon skeleton having two pentagons connected to highly symmetric five hexagons. Such unique structure brings above infrared mode. Advanced study on other molecules with such configuration is necessary to find more precise spectrum.

## 1. INTRODUCTION

The interstellar gas and dust show featured mid-infrared emission from 3 to 20µm. Discrete emission features at 3.3, 6.2, 7.6, 7.8, 8.6, 11.2, and 12.7µm are ubiquitous peaks observed at many astronomical objects (Ricca et al. 2012; Geballe et al. 1989; Verstraete et al. 1996; Moutou et al. 1999; Hony et al. 2001; Meeus et al. 2001; Peeters et al. 2002; Regan et al.2004; Engelbracht et al. 2006; Armus et al. 2007; Sloan et al. 2005; Smith et al. 2007; Sellgren et al. 2007). Current common understanding is that these astronomical spectra come from the vibrational modes of polycyclic aromatic hydrocarbon (PAH) molecules. Concerning PAH spectra, there are many experimental laboratory spectroscopy data (Szczepanski & Vala 1993a; Schlemmer et al. 1994; Moutou et al. 1996; Cook et al. 1998; Piest et al. 1999; Hudgins & Allamandola 1999a, 1999b; Oomens et al. 2001, 2003, 2011;Kim et al. 2001). and density functional theory (DFT) based theoretical analysis (de Frees et al. 1993; Langhoff 1996;Malloci et al. 2007; Pathak & Rastogi 2007; Bauschlicher et al.2008, 2009, 2010; Ricca et al. 2010, 2012).

The current central concept to understand these observed astronomical spectra is the decomposition method using many PAHs data (Boersma et al. 2013, 2014).

My question is that if there were some selection rule of the astronomical chemical evolution, there were limited numbers of survived PAHs. The aim of this paper is to find out a basic model molecule covering a wide range infrared spectrum (3~20µm). As the first trial, smallest and simplest coronene ($C_{24}H_{12}$) family was analyzed using DFT calculation. Ricca et al. (2012) made a remarkable study focusing on coronene family. They enhanced from $C_{24}H_{12}$ to $C_{384}H_{48}$, and had a success to reproduce observed spectra by the decomposition method using many coronene and ovalene family data.

This study intentionally stepped into void created molecules as like $C_{23}H_{12}$. Among several model molecules, a single void induced cation $C_{23}H_{12}^{++}$ brought successful spectrum. Calculated peaks were 3.2, 6.4, 7.6, 7.8, 8.6, 11.4, and 12.9µm, which were amazing coincidence with the observation.

## 2, CALCULATION METHOD

We have to obtain total energy, optimized atom configuration, and infrared vibrational mode frequency and strength depending on a respective given initial atomic configuration, charge and spin state Sz. Density functional theory (DFT) based hybrid B3LYP functional (Becke 1993; Stephans et al. 2009) was applied utilizing Gaussian09 package (Frisch et al. 2009) with an atomic orbital 6-31G basis set. The first step calculation is to obtain the self-consistent energy, optimized atomic configuration and spin density calculations. The required convergence on the root mean square density matrix was less than 10E-8 within 128 cycles. Using this result, atomic vibrational mode frequency and strength was calculated. Vibrational absorption strength is obtained as molar absorption coefficient ε (litter mol$^{-1}$ cm$^{-1}$).

Comparing such DFT based harmonic frequencies $N_{DFT}$(cm$^{-1}$) with experimental data, a single scale factor 0.958 was used for 3~20 μm range (Ricca et al. 2012).

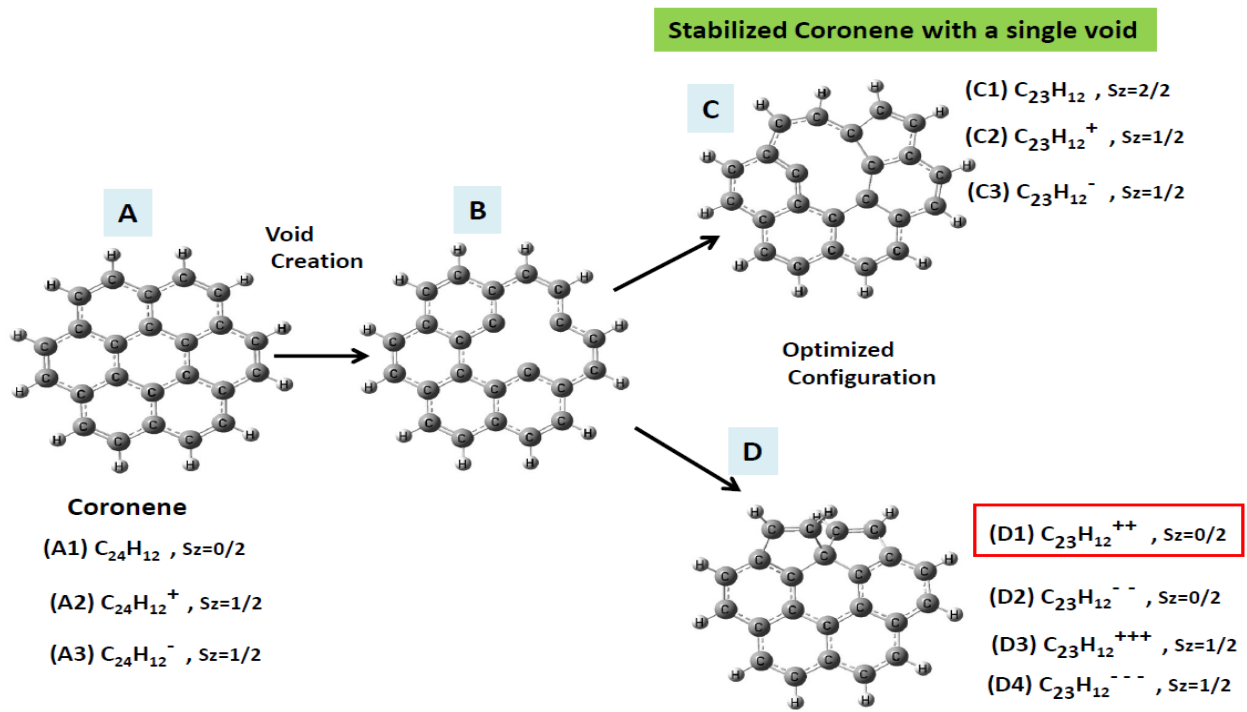

**Figure 1.** Optimized atomic configuration of coronene family classified by void, charge and spin

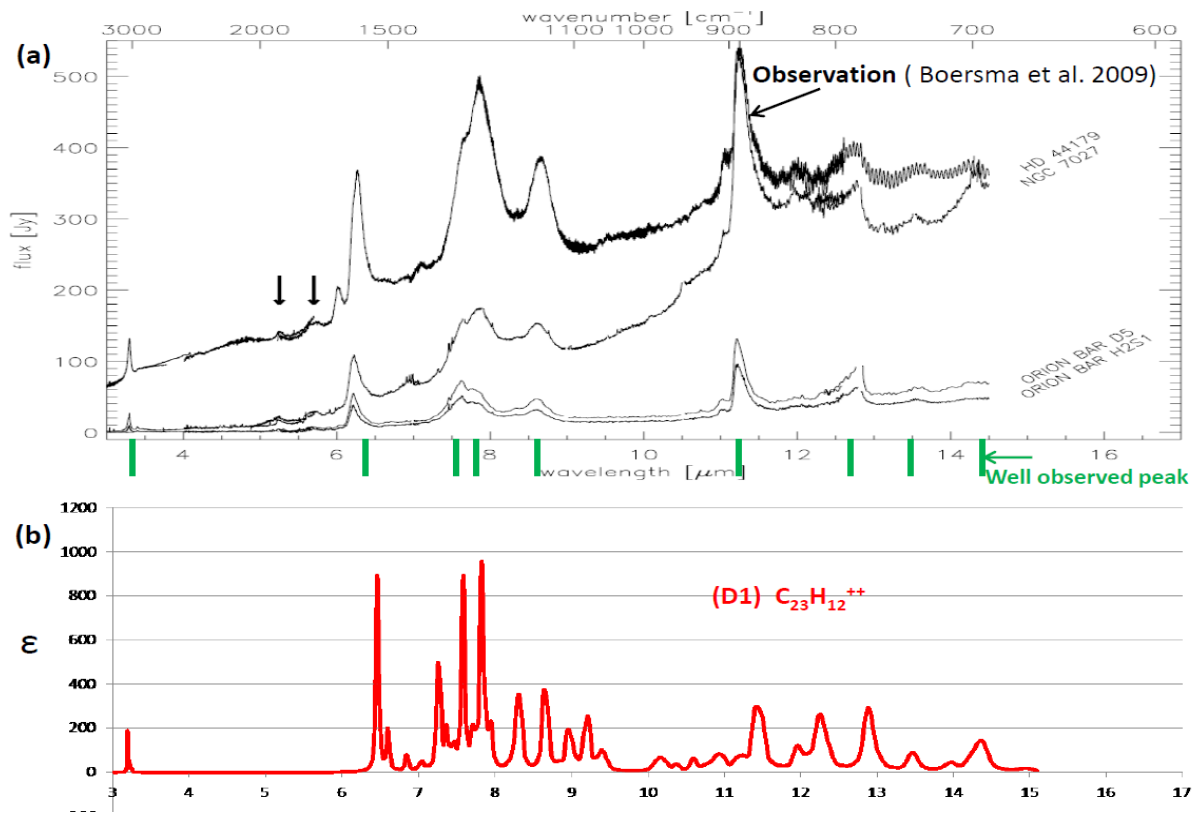

**Figure 2**. Astronomically observed infrared spectra (upper figure) of four sources, and DFT based calculated spectra of $C_{23}H_{12}^{++}$ single molecule (lower figure) covering wavelength from 3 to 15 $\mu$ m

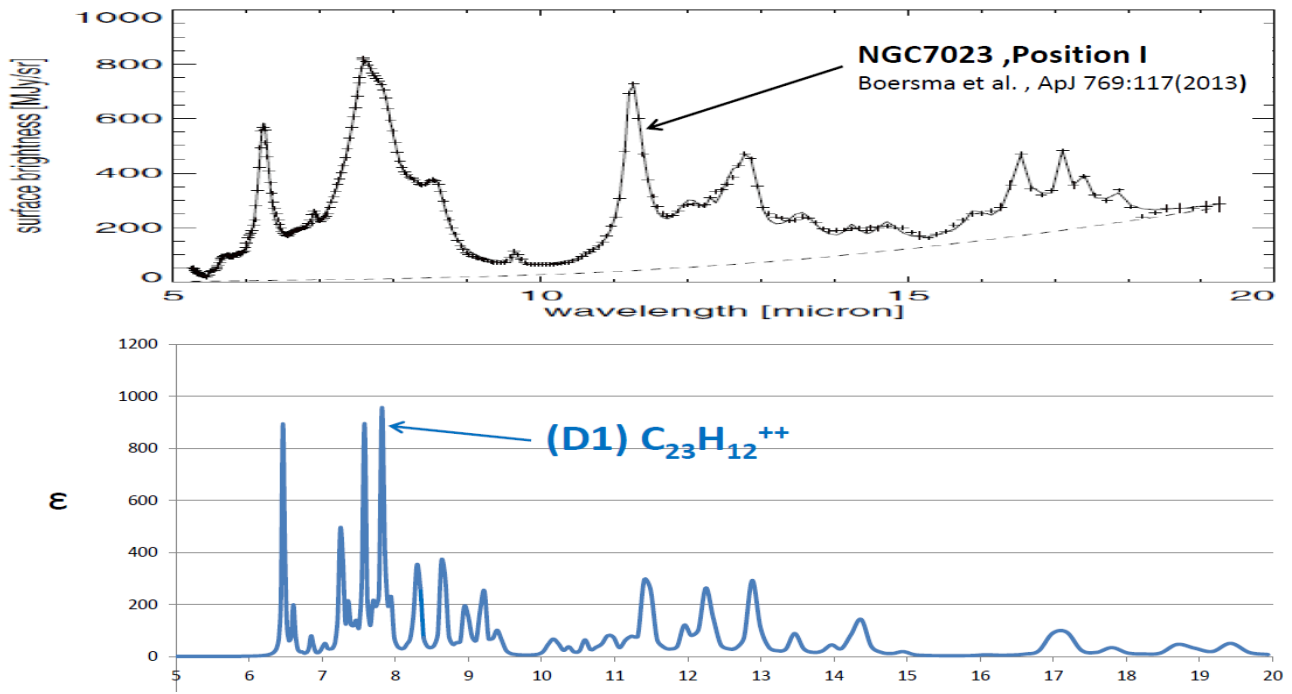

Figure 3. Comparison of photon dominated NGC7023 region spectra with calculated $C_{23}H_{12}^{++}$ one.

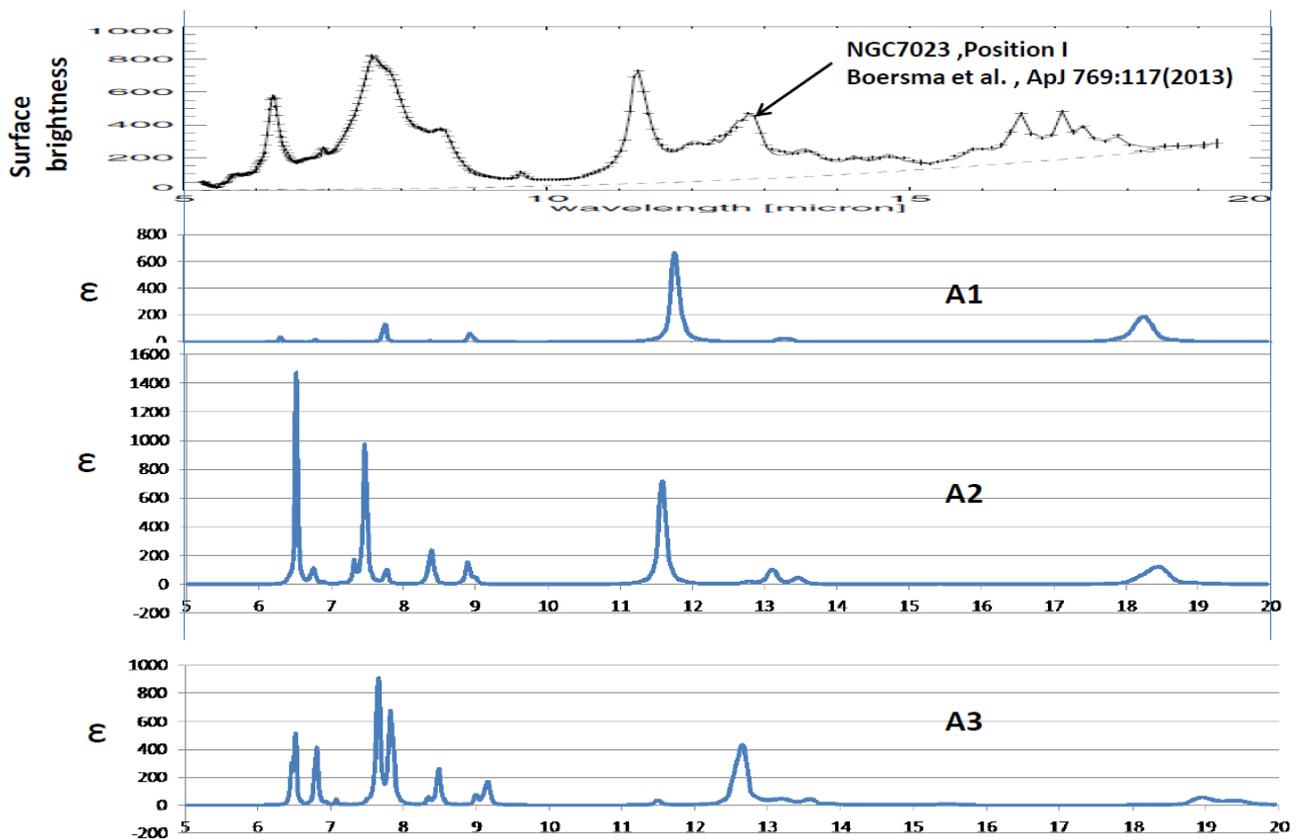

Figure 4.  Calculated spectra of coronene $C_{24}H_{12}$ (A1), $C_{24}H_{12}^{+}$ (A2), and $C_{24}H_{12}^{-}$ (A3)

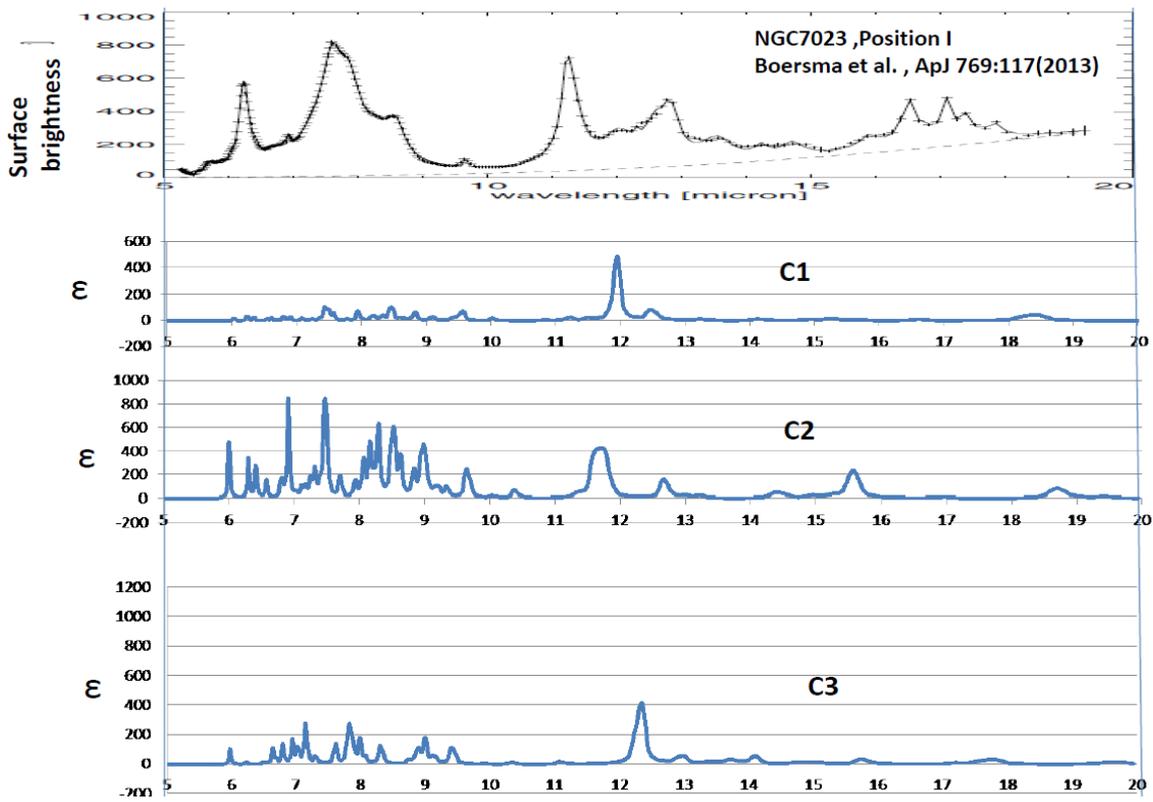

**Figure 5.** Calculated spectra of single void coronene $C_{23}H_{12}$ (C1), $C_{23}H_{12}^{+}$ (C2), and $C_{23}H_{12}^{-}$ (C3)

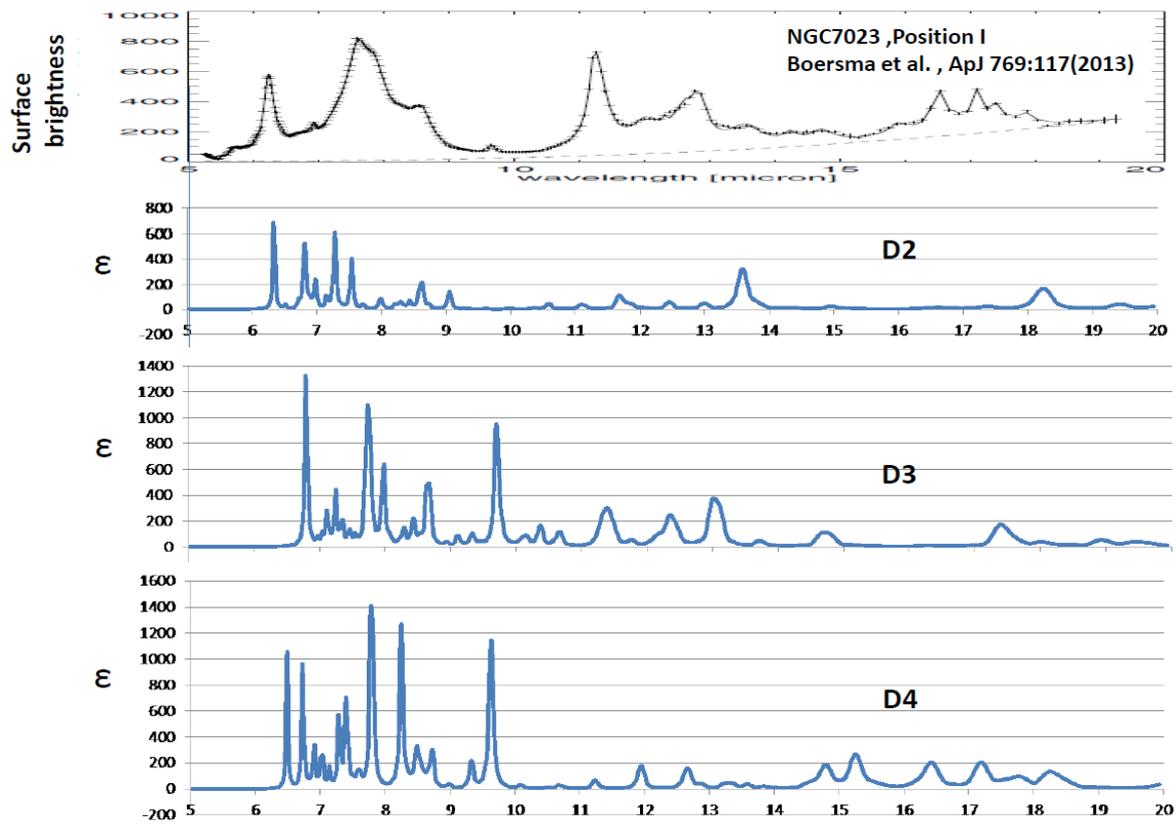

**Figure 6.** Calculated spectra of single void coronene $C_{23}H_{12}^{--}$ (D2), $C_{23}H_{12}^{++}$ (D3), and $C_{23}H_{12}^{---}$ (D4)

Computed spectra are for absorption. However we need astronomical PAHs emission spectra. As noted details by Ricca et al. (Ricca et al. 2012), we should consider photon absorbed emission and apply red shift by 15 cm$^{-1}$. Final calculated emission frequency (wave number) N (cm$^{-1}$) is,

$N(cm^{-1}) = N_{DFT}(cm^{-1}) \times 0.958 - 15 (cm^{-1})$

Also, wavelength λ is,

$\lambda(\mu m) = 10000/N(cm^{-1})$

Emission strength is assumed to be proportional to DFT calculated ε. For figuring λ vs ε, frequency distribution method included in windows excel function was applied. Half width of λ was 0.1µm.

### 3, MODEL CORONENE FAMILY

Model coronene family was illustrated in Figure 1. Starting model was basic coronene (group A) of neutral $C_{24}H_{12}$ (A1), cation $C_{24}H_{12}^+$ (A2), and anion $C_{24}H_{12}^-$ (A3). For optimizing atomic configuration and obtaining converged energy, we need to give a precise spin parameter Sz. A1 is non-magnetic with Sz=0/2. Whereas, A2 and A3 have an extra charge to give Sz=1/2. In this study, carbon single void was supposed as shown in B. In astronomical gas and dust surrounding an extinct star and/or new born star, high energy photon and particles (electron, proton, carbon, Si, Mg etc.) may attack group A molecules to make unstable void molecule B. Optimized atomic configuration was classified into two types as C and D. Type C include neutral $C_{23}H_{12}$ (C1) with Sz=2/2 (Ota; 2011, 2014), mono-cation $C_{23}H_{12}^+$ (C2) Sz=1/2, and mono-anion $C_{23}H_{12}^-$ (C3) Sz=1/2. They are all magnetic. Optimized atomic configuration was asymmetric. Inside of the molecule, there was one carbon pentagon connected with one hexagon. This structure was brought by the Jahn-Teller distortion effect (Ota; 2014). It should be noted that in case of double charged ($C_{23}H_{12}^{++}$ (D1), $C_{23}H_{12}^{--}$ (D2)) and triple charged ($C_{23}H_{12}^{+++}$ (D3), $C_{23}H_{12}^{---}$ (D4)) molecules, structure recovers high symmetry as shown in D. There occurs serious Jahn-Teller distortion on carbon skeleton bringing two pentagons connected with highly symmetrical five hexagons. Magnetism of D1 and D2 is nonmagnetic with Sz=0/2 by binary electrons spin canceling. Side view of D was like a character "Y". Two pentagons declined each other with an angle of 95 degree.
.

### 4, MID-IR SPECTRA OF $C_{23}H_{12}^{++}$

Among ten candidates, most interesting spectra was obtained in $C_{23}H_{12}^{++}$ (D1) as shown in Figure 2. Upper figure in Figure 2 was astronomical observation data compiled by Boersma et al. (Boersma et al. 2009). The four sources are HD44179, NGC7027, Orion ionization ridge D5 and H2S1. Those are thought to be rich PAH region. Infrared high peaks are 3.3, 6.3, 7.6, 7.8, 8.6, 11.2, 12.7, 13.5 and 14.3µm as marked by green line. Whereas, calculated peaks (red curve at lower figure) are **3.2**, **6.4**, 7.2, **7.6**, **7.8**, 8.3, **8.6**, 8.9, 9.2, **11.4**, **12.9**, **13.5**, and **14.4**µm. Identical wavelength within 0.2µm was marked by block character. This is an amazing coincidence by a single molecule, not using the decomposition method. While, calculation could not reproduce small peaks as like 5.25 and 5.7µm as discussed by Boersma et al. (Boersma et al. 2009). Calculated 7.2µm peak was observed only at HD44179 spectra. Calculated 8.3, 8.9, and 9.2µm peaks could not find in those four sources. Another example was compared in Figure 3. Cross marked curve is the photon dominated region spectra of NGC7023 (Boersma et al. 2013, 2014). Calculated peaks also reproduced well again observed one. We can believe that the spectra of $C_{23}H_{12}^{++}$ (D1) tell us some important message to find out how PAH created and survived in harsh environment of interstellar space. Also, it is very interesting that pentagon-hexagon coupled skeleton is similar with biological basic elements on the earth.

### 5, OTHER CORONENE FAMILY

Calculated spectra of Group A was shown in Figure 4. Neutral coronene $C_{24}H_{12}$ (A1) has small peaks at 6.3, 7.8, 8.9µm, which are very well coincide with previous calculated results (Ricca et al. 2012). Also, cation $C_{24}H_{12}^+$ (A2) and anion $C_{24}H_{12}^-$ (A3) show a similar tendency. However, unfortunately every single molecule could not cover a wide range observed spectra. Results of group C was illustrated in Figure 5. Calculated spectra were very different with observed one, and very complicated, which may reflect an asymmetric molecule structure as shown in C of Figure 1. Double and triple charged group D results were obtained in Figure 6 except D1. Tri-cation $C_{23}H_{12}^{+++}$ (D3) show some similarity with observed peaks, but not so good compared with di-cation $C_{23}H_{12}^{++}$ (D1).

### 6, DISCUSSION

This paper could successfully find out one peculiar molecule $C_{23}H_{12}^{++}$ covering wide range (3~20µm) astronomically observed infrared spectra.

Such calculated spectra come from unique molecular configuration, that is, carbon two pentagons have connected highly symmetrical five hexagons. Therefore, larger molecules having such skeleton may have a capability reproducing observed spectra. Candidates are circum coronene, circum circum coronene, ovalen family and more carbon number PAHs.

It is essential to study the principle how particular PAHs were selected in harsh space conditions, which mean the chemical evolution process in the universe.

Carbon based pentagon-hexagon coupled skeleton is familiar to biological organic components. Missing link between astronomical PAHs and biological components will be final question.

## 7, CONCLUSION

In order to find out a single molecule reproducing astronomically observed infrared spectra, not depending on the decomposition method using many PAHs, coronene family model molecules including voids and charges have been computed using density functional theory. Among ten model molecules, a single void induced di-cation $C_{23}H_{12}^{++}$ have successfully reproduced a wide range spectra (3~20 µm) of typical interstellar gas and dust spectra. Well known astronomically observed emission peaks of 3.3, 6.3, 7.6, 7.8, 8.6, 11.2, 12.7, 13.5 and 14.3µm were successfully reproduced by a single molecule $C_{23}H_{12}^{++}$, which calculated peaks were 3.2, 6.4, 7.6, 7.8, 8.6, 11.4, 12.9, 13.5, and 14.4µm. Such coincidence suggested that some astronomical chemical evolution in interstellar space may lead a particular selection among many polycyclic aromatic hydrocarbon candidates. Optimized molecular structure of $C_{23}H_{12}^{++}$ show carbon skeleton with two pentagons connected to highly symmetrical five hexagons. Such unique structure was thought to bring above infrared mode. Larger molecules including such basic molecular structure will be expected to show similar spectra.